\documentclass[12pt]{iopart} 
 
\usepackage{amsfonts}
\usepackage{amssymb}
\usepackage{bm}
\usepackage[dvips]{color}
\usepackage{graphicx}
\usepackage{subfigure}

\newcommand{\hide}[1]{}

\begin{document}
 
\newcommand{\yjc}[1]{\textcolor{blue}{#1}} 
%%%%%%%%%%%%%%%%%%%%%%%%%%%%%%%%%%%%%%%%%%%%%%%%%%%%%%%%%%%%%%%%%
%%%%%%%%%%%%%%%%%%%%%%%%%%%%%%%%%%%%%%%%%%%%%%%%%%%%%%%%%%%%%%%%%
 
\title{Tuning the adiabaticity of spin dynamics in diamond nitrogen vacancy centers}
 
\author{Y. B. Band$^{1}$ and Y. Japha$^{2}$}
\address{
$^{1}$Department of Chemistry, Department of Physics, and the Ilse Katz Center for Nano-Science,
Ben-Gurion University, Beer-Sheva 84105, Israel\\
$^{2}$Department of Physics, and the Ilse Katz Center for Nano-Science,
Ben-Gurion University, Beer-Sheva 84105, Israel}
% \email{band@bgu.ac.il}
% \date{\today}
 
\begin{abstract}
We study the spin dynamics of diamond nitrogen vacancy (NV) centers in an oscillating  magnetic field along the symmetry axis of the NV in the presence of transverse magnetic fields. It is well-known that the coupling between the otherwise degenerate Zeeman levels $|M_S=\pm1\rangle$ due to strain and electric fields is responsible for a Landau-Zener process near the pseudo-crossing of the adiabatic energy levels when the axial component of the oscillating magnetic field changes sign. We derive an effective two-level Hamiltonian for the NV system that includes coupling between the two levels via virtual transitions into the third far-detuned level $|M_S=0\rangle$ induced by transverse magnetic fields. This coupling adds to the coupling due to strain and electric fields, with a phase that depends on the direction of the transverse field in the plane perpendicular to the NV axis. Hence, the {\em total coupling}  of the Zeeman levels can be tuned to control the adiabaticity of spin dynamics by fully or partially compensating the effect of the strain and electric fields, or by enhancing it.  Moreover, by varying the strength and direction of the transverse magnetic fields, one can determine the strength and direction of the local strain and electric fields at the position of the NV center, and even the {\em external} stress and electric field.  The nuclear spin hyperfine interaction is shown to introduce a nuclear spin dependent offset of the axial magnetic field for which the pseudo-crossing occurs, while the adiabaticity remains unaffected by the nuclear spin.  If the NV center is coupled to the environment, modeled by a bath with a Gaussian white noise spectrum, as appropriate for NVs near the diamond surface, then the spin dynamics is accompanied by relaxation of the Zeeman level populations and decoherence with a non-monotonic decrease of the purity of the system.
The results presented here have important impact for metrology with NV centers, quantum control of spin systems in solids and coupled dynamics of spin and rotations in levitated nano-objects in the presence of magnetic fields.
\end{abstract}


\maketitle
 
\section{Introduction}   \label{sec:Intro}
 
Negatively charged nitrogen-vacancy (NV) color centers in diamond \cite{Doherty_13} have been proposed as candidates for sensors of various physical events or environmental changes, including biological fluorescent markers~\cite{Sage_13}, qubits~\cite{Gruber_97,Oort_88, Jiang_09}, magnetic field and electric field sensors \cite{Balasubramanian_09,Dolde_11}, stress sensors \cite{Kehayias}, temperature sensors~\cite{Neumann_13,Kucsko_13} and inertial sensors~\cite{Ledbetter_12}. 
The peculiar properties of a NV center in diamond stem from the fact that it is a hybrid creature that combines a solid object (the diamond crystal) with a system consisting of a single spin with discrete energy levels like an atom or a molecule. Unlike atoms, NV centers do not suffer from the broadening effects of atomic systems such as Doppler broadening due to translational motion or power broadening due to collisions. And, unlike molecules, they do not have the broad spectrum of rotations or vibrations, so that at room temperature NV centers show a very precise spectrum of discrete levels. This is why NV centers are ideal for studying fundamental quantum processes, and for ultra-precise sensing. In addition, a levitated, cooled and trapped micro- or nano-diamond with a single or many spins is a good candidate for implementing a fully quantum object with a large mass and coupling between its internal spin and its external motion for studying the transition between quantum and  classical mechanics~\cite{Hsu_2016, Delord_2020}.  Manipulation of such an object with external electric, magnetic or optical fields requires a thorough study of the response of a NV center to such external fields.

The ground electronic state of the NV center is a triplet ($S=1$) with a large splitting [${\cal D}  = h \times 2.87$ GHz, where $h$ is Planck's constant] between the magnetically sensitive Zeeman states $|M_S=\pm1\rangle$ and the magnetically insensitive state $|M_S=0\rangle$. It follows that relatively weak magnetic fields (of less than hundreds of Gauss) do not induce transitions between the Zeeman states. However, the degeneracy of the states $|M_S = \pm 1 \rangle$ states at zero magnetic field is lifted by coupling between them due to strain and local electric fields such that the eigenstates of the two-level system create two energy branches avoided crossing between them. The spin dynamics resulting when temporally changing magnetic fields are present is that of Landau-Zener (LZ) transitions when the change is fast, and adiabatic dynamics,  wherein the population stays at one of these energy branches, when the change of the magnetic field is slow. 

The two-level Landau-Zener (LZ) problem \cite{Landau, Zener, Stueckelberg, Majorana} in a system with avoided level crossing was first suggested in 1932 to theoretically model molecular pre-dissociation. LZ dynamics have been observed in many systems, including solid-state systems, such as superconductor two-level systems~\cite{Shevchenko_2010} and systems having paramagnetic defects in semiconductors \cite{Awschalom_14, Miao_Awschalom_20}.  Moreover, LZ dynamics was observed in diamond NV centers subjected to microwave or radio-frequency magnetic fields~\cite{Fuchs_09, Fuchs_11, Huang_Du_11, Dmitriev_19}, and were used in quantum memory elements \cite{Fuchs_11}.  In this context, when the axial magnetic field component is scanned along the avoided crossing the Landau-Zener transition probability is given by 
 (see appendix~\ref{app:adiabaticity}) 
\begin{equation} \label{eq:P_LZ_NV}
P_{LZ}=\exp\left(-\frac{\pi|\epsilon|^2}{\hbar\mu|\dot{B}_{\parallel}|}\right), 
\end{equation}
where $\epsilon$ is the coupling (interaction) between the $|M_S=\pm1\rangle$ states, $\mu$ is the magnetic moment of the NV center and $\dot{B}_{\parallel}$  is the rate of change of the axial magnetic field when it crosses through $B_{\parallel}=0$. 

The dynamics of the spin across such a pseudo-crossing is the subject of this paper. In particular, we address the question of whether the spin projection along the axis follows the projection of the magnetic field (adiabatic dynamics) or stays aligned along the original direction while the magnetic field flips its projection (non-adiabatic dynamics).  Landau-Zener transitions, and in particular adiabaticity of spin-1 (three-level) dynamics in the presence of driving fields, was studied in previous theoretical work as a demonstration of general principles of quantum control~\cite{Kenmoe_16, Xu_18, Wu_19, Xu_19}.  Here we examine the practical case of diamond NV centers, where an intrinsic energy gap due to the presence of strain and electric fields results in a pseudo-crossing even in the absence of transverse magnetic fields. We show that a transverse magnetic field can be used to control the final spin state of the NV due to changing the adiabaticity of the spin dynamics. 
This may inspire new methods for manipulating the spin states without involving optical or even microwave frequencies, as those used in coherent manipulation methods such as stimulated Raman transition or adiabatic passage~\cite{Bohm_2021}. Moreover, using a transverse magnetic field, one can measure the local strain and electric field strengths and directions  at the NV center position. This might even be used as a sensor to measure external strains and electric fields.

The outline of this paper is as follows: In Sec.~\ref{sec:tune_adiab} we study the effect of transverse magnetic fields on the adiabaticity of spin dynamics. We then investigate the effect of the hyperfine interactions on the dynamics in Sec.~\ref{sec:hyperfine}.  The nuclear degrees of freedom do not significantly change the nature of the dynamics but modify it in a way that allows its description on the basis of the dynamics of the electronic system that we discussed above.  Section \ref{sec:decoherence} considers the dynamics including decoherence due to isotropic white noise using a formalism developed in Ref.~\cite{Band_3_level_19} which considered the three-level LZ problem for open system cases where interaction with an environment is present. When the environment can be modeled as white Gaussian noise, the system can be treated using a master equation of the form $i \, \partial \rho/\partial t = [H(t), \rho(t)] - \Gamma \rho(t)$, where $\rho$ is the density matrix and $\Gamma$ is the Lindblad decay operator.  Finally, Sec.~\ref{sec:S&C} presents a summary and conclusions and briefly discusses some implications of this work.  Three appendices clarify some aspects of the discussion in the main text.

\section{Effect of Transverse Magnetic Field on Adiabaticity} 
\label{sec:tune_adiab}

In this section we analyze the spin dynamics of the electronic ground state of a NV center in the presence of time-dependent magnetic fields and show that although transverse magnetic field components cannot directly induce spin rotation, they can still significantly affect the adiabaticity of spin dynamics along the axis. In particular, we show that it is possible to tune adiabaticity (or, more precisely, non-adiabaticity) with a transverse magnetic field.  The quantity which is responsible for the pseudo-crossing of the adiabatic eigenvalues is the `strain coefficient' $\epsilon$  (see below).  We show that it is possible to counterbalance the strain (and electric field) effects with a transverse magnetic field, and even cause the effective interaction between the $M_S = \pm 1$ levels to vanish, by tuning the strength and direction of the transverse magnetic field.

The electronic ground state of a NV center is a triplet (spin $S=1$) with a large splitting ${\cal D} =h\times 2.87$ GHz between the level with projection of the angular momentum along the axis parallel to the nitrogen-vacancy vector ($\hat{\bf z}$ axis),  $M_S = 0$, and the $M_S = \pm 1$ levels. 
The electronic Hamiltonian in the presence of an arbitrary magnetic field ${\bf B}$ can be written as \cite{Doherty_13}
\begin{eqnarray}  \label{eq:H_nv}
    H(t)&=& {\cal D} \left(S_z^2 -\frac{2}{3} {\bf 1}\right)  + \mu \, {\bf S} \cdot {\bf B}(t) + H_{es},
\end{eqnarray}
where $\mu=h\times 2.8$\,MHz/G is the magnetic moment of the NV and the spin angular momentum components $S_i$ ($i = x, y$ and $z$) are represented by spin-1 3$\times$3 matrices \cite{Band_Avishai_1}. $H_{es}$ is the interaction Hamiltonian of the NV center with strain and local electric fields  (the subscripts $e$ and $s$ stand for electric and strain).  It removes the axial symmetry about the $\hat{\bf z}$ axis, couples the $M_S = \pm 1$ levels and removes their degeneracy. It has the form
\begin{equation} \label{eq:Hstrain} 
H_{es}=\epsilon_{es}^x(S_x^2-S_y^2)+\epsilon_{es}^y(S_xS_y+S_yS_x).
\end{equation}
Here $\epsilon_{es}^j = d^\perp [\delta_j + E_j]$ ($j=x,y$),  where $E_x$ and $E_y$ are the transverse electric field components at the position of the NV, $\delta_x$ and $\delta_x$ are the strain field components, and $d^\perp$ is the transverse component of the ground state electric dipole moment \cite{Doherty_13}.  Experimental measurement \cite{Oort} of the transverse electric dipole moment showed that $d^\perp = 17 \pm 3$ Hz cm/V [and the longitudinal electric dipole moment $d^\parallel$ is more than an order of magnitude smaller than $d^\perp$, hence it is neglected here].  The intrinsic strain coefficients $\epsilon_{es}^x,\epsilon_{es}^y$ can vary considerably from sample to sample because of variations in the local electric and strain fields, and is typically in the range of several MHz.  
Note that the form of $H_{es}$ in Eq.~(\ref{eq:Hstrain}) is invariant under the choice of axes in the plane perpendicular to the $\hat{z}$ direction.

In a matrix form, Hamiltonian~(\ref{eq:H_nv}) is given by,
\begin{equation}    \label{Eq:H_nv_general}
H=\left(\!\!
\begin{array}{ccc}
    \mu B_z & \mu\frac{B_x-iB_y}{\sqrt{2}} & \epsilon_{es}^*  \\
  \mu\frac{B_x+iB_y}{\sqrt{2}} & -{\cal D} & \mu\frac{B_x-iB_y}{\sqrt{2}} \\
  \epsilon_{es}  & \mu\frac{B_x+iB_y}{\sqrt{2}} & -\mu B_z\end{array} \!\!\right)
+\frac13{\cal D} \, {\bf 1}.
\end{equation}
where $\epsilon_{es}\equiv \epsilon_{es}^x+i\epsilon_{es}^y=|\epsilon_{es}|e^{i\phi_{es}}$ is complex and the phase $\phi_{es}$ in the complex plane corresponds to the angle of the strain+electric field at the NV center in the transverse plane. 

\begin{figure}% [ht]
\centering
\includegraphics[width=0.9\linewidth]{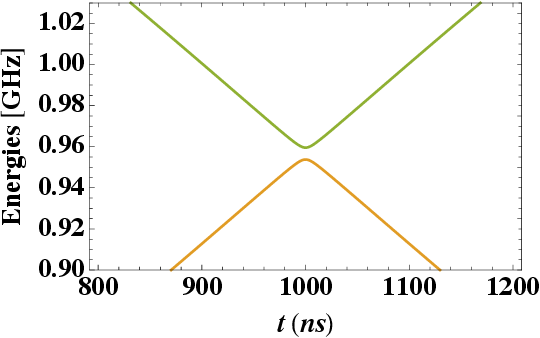}
\caption{Adiabatic energy eigenvalues [Eq.~(\ref{eq:eig})] for the case of a magnetic field ${\bf B}(t)=\hat{\bf z}B_0\cos\omega t$. The energy scale is relative to the center of the ground state triplet energy. 
The parameters are $\epsilon =\epsilon_{es}= h\times 2.87$\,MHz, $\mu B_0 = h\times 28$\,MHz (corresponding to a magnetic field of 10 Gauss) and angular frequency $\omega = 2 \pi \times 0.25$ MHz. The adiabaticity parameter is $\beta=\hbar\omega\mu B_0/\pi\epsilon^2=0.26\ll 1$. The energy eigenvalues are shown only in the range close to the pseudo-crossing at $t=\pi/2\omega=1000$\,nm. 
}
\label{Fig_1}
\end{figure}

The spin of the NV center is sensitive mostly to the magnetic field component along the axis of the NV center (although the effects of the transverse magnetic field components can become important, as shown in this section).  The spin dynamics is constrained to the NV axis due to the strong breaking of the spherical symmetry, which is represented by the zero-field splitting parameter ${\cal D}$ of the $|M_S=0\rangle\equiv |0\rangle$ state from the $|M_S=\pm 1\rangle\equiv |\pm1\rangle$ states. Due to the large splitting ${\cal D}$ we can eliminate the $M_S=0$ level due to its high energy relative to the other energy scales involved. Using the Feshbach formalism~\cite{Feshbach}, we apply the projection operators $P=|1\rangle\langle 1|+|-1\rangle\langle -1|$ and $Q=1-P=|0\rangle\langle 0|$ to separate between the states $|\pm 1\rangle$ and the level $|0\rangle$ and derive an effective $2\times 2$ Hamiltonian for the levels $|\pm 1\rangle$ that is valid whenever the energies and rates of change involved in their interaction with the magnetic field are much smaller than the zero-field splitting ${\cal D}$. When the energies of the part of Hilbert space that is of interest (in our case the states $|\pm1\rangle$) are much smaller than their energy separation from the other part of the Hilber space (in our case the state $|0\rangle$), we show in \ref{app:Heff} that the effective Hamiltonian for the states of interest has the form
\begin{equation} H_{\rm eff}=PHP-PHQ\frac{1}{QHQ}QHP, 
\label{eq:Heff_general}
\end{equation}
For the Hamiltonian of the NV center in Eq.~(\ref{Eq:H_nv_general}) $PHP$ is the block contained in the first and third rows and columns, while $PHQ=(QHP)^{\dag}=\frac{\mu}{\sqrt{2}}[(B_x-iB_y)||+1\rangle\langle 0|+(B_x+iB_y)|-1\rangle\langle 0|]$ and $QHQ=-{\cal D}|0\rangle\langle 0|$. 
The second term on the right-hand-side of Eq.~(\ref{eq:Heff_general}) represents the effect of virtual transitions from the states $|\pm1\rangle$ to the state $|0\rangle$ on the dynamics of the states $|\pm1\rangle$, while the far detuned state $|0\rangle$ is eliminated from the explicit dynamics. In our case the effective Hamiltonian of Eq.~(\ref{eq:Heff_general}) has the explicit form
\begin{equation}  \label{eq:Heff}
H_{\rm eff}=\left(\!\!
\begin{array}{cc}
  \mu B_z & \epsilon({\bf B}_{\perp})^* \\
  \epsilon({\bf B}_{\perp}) & -\mu B_z\end{array} \!\!\right) +\eta B_{\perp}^2\hat{1}, 
\end{equation}
where ${\bf B}_{\perp}=(B_x,B_y)=B_{\perp}(\cos\phi,\sin\phi)$ is the projection of the magnetic field into the transverse $x$-$y$ plane, having an angle $\phi$ in the plane, 
$\eta=\mu^2/2{\cal D}$, and
\begin{equation} \label{eq:epsilon_B}
\epsilon({\bf B}_{\perp})=\epsilon_{es}+\eta B_{\perp}^2e^{2i\phi}. 
\end{equation}
The eigenvalues of this effective Hamiltonian are
\begin{equation} \label{eq:eig}
E_{\pm}({\bf B})= \eta B_{\perp}^2\pm \sqrt{\mu^2B_z^2  + 
|\epsilon({\bf B}_{\perp})|^2}. 
\end{equation}
The effective Hamiltonian~(\ref{eq:Heff}) and the energy eigenvalues in Eq.~(\ref{eq:eig}) are valid when $\mu |{\bf B}|\ll {\cal D}$. As shown in an improved approximation for the energy eigenvalues, which extends the validity range to magnetic fields values that are only somewhat smaller than ${\cal D}/\mu$, is obtained by replacing ${\cal D}$ in the expression for $\eta$ by ${\cal D}-E_{\pm}$, where $E_{\pm}$ is taken from the lower order approximation in Eq.~(\ref{eq:eig}) with $\eta=\mu^2/2{\cal D}$ (see \ref{app:Heff}). This improved approximation is not needed here because we will only be interested in the adiabaticity when the magnitudes of the adiabatic eigenvalues are near their minimum.

If the system is initially in one of the energy eigenstates whose eigenvalues are given in Eq.~(\ref{eq:eig}), then, when the magnetic field changes slowly enough, the system will stay in the same adiabatic eigenstate, whose energy is given by $E_+({\bf B}(t))$ or $E_{-}({\bf B}(t))$. Each of these adiabatic eigenstates is a spin state where the spin direction with respect to the projection of the spin on the NV axis is conserved for all values of ${\bf B}(t)$. 
When the axial component $B_z$ of the magnetic field is swept through $B_z=0$, the probability for LZ transition is given by Eq.~(\ref{eq:P_LZ_NV}) and therefore the adiabaticity of the dynamics is determined by the dimensionless parameter 
\begin{equation} \label{eq:beta}
\beta=\frac{\hbar\mu}{\pi|\epsilon|^2}\left|\dot{B}_z\right|_{B_z=0}, 
\end{equation}
where $\epsilon$ is given in Eq.~(\ref{eq:epsilon_B}).  If $\beta\ll 1$ the dynamics is adiabatic and the system stays in the adiabatic state where it started, {\it i.e.}, the spin state stays in  the same direction with respect to the magnetic field projection on the $\hat{z}$ axis as it was before the splitting. In contrast, when $\beta\gg 1$ the system changes its direction with respect to the magnetic field projection: a LZ transition occurs and the system transforms into the other energy branch.  As the effective coupling strength $|\epsilon|$ depends on the transverse components of the magnetic field, this part of the field may be used to tune the adiabaticity of the process when $B_z$ is swept throuth $B_z=0$. 

To demonstrate tuning of the adiabaticity let us now consider an oscillating magnetic field $B_z(t)=B_0\cos(\omega t)$ in the longitudinal direction. Figure~\ref{Fig_1} presents the adiabatic energy eigenvalues [eigenvalues of the instantaneous Hamiltonian $H(t)$] for $B_{\perp}=0$. In this case the off-diagonal coupling matrix element $\epsilon({\bf B}_{\perp}=0)$ is equal to the intrinsic value $\epsilon_{es}$, which we take to be real and have a typical value of $\epsilon_{es}=10^{-3}{\cal D}= h\times 2.87$\,MHz. For this value of the splitting between the two branches of the energy eigenvalues and the frequency $\omega/2\pi=0.25$\,MHz the adiabaticity factor for LZ transitions in Eq.~(\ref{eq:beta}) is $\beta=0.26 \ll 1$ so that the probability for a LZ transition is very low and the dynamics is adiabatic (see numerical demonstration in \ref{app:longi}). 

\begin{figure} %[ht]
\centering
\includegraphics[width=0.8\linewidth]{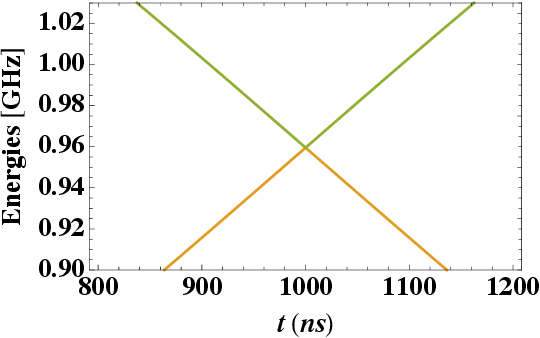}
\caption{Adiabatic eigenvalues of the three-level Hamiltonian in Eq.~(\ref{Eq:H_nv_general}) as a function of time for $B_z(t)=B_0\cos\omega t$. 
Here $\epsilon_{es}$ is real and $B_y = \sqrt{2{\cal D}\epsilon_{es}}/\mu\equiv B_c$, such that $\epsilon({\bf B}_{\perp})=0$ in Eq.~(\ref{eq:eig}) (other parameters as in Fig.~\ref{Fig_1}).  The adiabatic eigenvalues cross, rather than pseudo-cross, hence the dynamics is expected to be completely non-adiabatic.  
The third adiabatic eigenvalue with $M_S = 0$ is a straight horizontal line at energy $-2.87$\,GHz$-2\eta B_y^2/h$ (where the second term is $\approx 5.74$\,MHz, and is due to the second-order interaction with the transverse field), and lies well outside of the plot range.}
\label{Fig_non-adiab}
\end{figure}

\begin{figure} %[ht]
\centering
\includegraphics[width=0.9\linewidth]{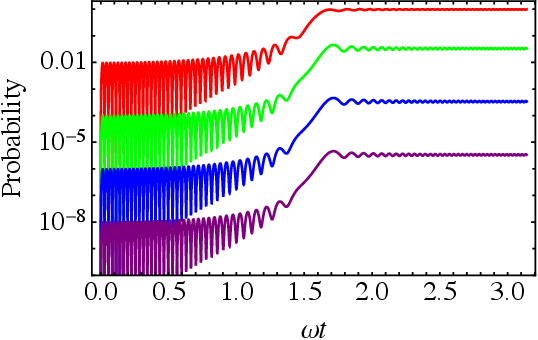}
\caption{Non-adiabatic dynamics with the population initially in the $M_S = 1$ level and $B_y = 0, 0.9487 \, B_c, 0.995 \, B_c, 0.9995\,B_c$ corresponding to $1-\eta B_y^2/\epsilon=1,\, 0.1,\,0.01$ and $0.001$. The other parameters are identical to those used to obtain Fig.~\ref{Fig_1}.  The figure plots the probability $P_{M_S = 1}(t)$, on a log scale, versus $\omega t$.}
\label{Fig_non-adiab-pop}
\end{figure}

The adiabaticity can be turned off so that LZ transitions have a 100\% probability, if the  effective splitting between the two states at $B_z=0$ vanishes. This can be achieved by setting $\eta B_{\perp}^2 = |\epsilon_{es}|$ and the direction of the transverse field to be perpendicular to the angle of the combined strain and electric fields, such that $2\phi=\pi$ for real $\epsilon_{es}$.  In the example below we use a real $\epsilon_{es}$ with $\epsilon_{es}=h\times 2.87$\,MHz and $B_{\perp}=B_y$, hence $\eta=h\times 2.8^2\,($MHz/G$)^2/(2\cdot 2.87$\,GHz$)=h \times 1.35$\,kHz/G$^2$. It follows that the specific magnetic field at which the levels become degenerate at $B_z=0$ is $B_{\perp}=\sqrt{|\epsilon_{es}|/\eta}=\sqrt{2{\cal D}|\epsilon_{es}|}/\mu\equiv B_c$, where for our choice of $\epsilon_{es}$, $B_c \approx 46$ G.  
Figure \ref{Fig_non-adiab} shows the adiabatic eigenvalues for a magnetic field with components $B_z(t) = B_0 \cos(\omega t)$ as in Fig.~\ref{Fig_1} and a static transverse field $B_y = B_c$. Clearly, the eigenvalues cross, rather than pseudo-cross.  We numerically solved the time-dependent Schr\"odinger equation with the effective Hamiltonian~(\ref{eq:Heff}), taking the initial state to be $M_S=1$ and all the parameters as in Fig.~\ref{Fig_1} except for the additional transverse magnetic field.  The results are shown in Fig.~\ref{Fig_non-adiab-pop}.  Clearly, coupling is turned off and non-adiabaticity ensues (e.g., for $B_y = 0.9995\,B_c$, the final probability to stay in the adiabatic level is $\approx 3.6 \times 10^{-6}$).  A full calculation with 9-levels including hyperfine interaction (see Sec.~\ref{sec:hyperfine}) is shown in Fig.~\ref{Fig_non-adiab-pop-9-level} and confirms this behavior.  The probability of staying in the adiabatic state (i.e., to be in the $M_S=-1$ manifold at the final time) is still very small even when the hyperfine interactions are taken into account.  Note the small probability for building up population in the $M_S = 0$ states during intermediate times, which do {\em not} fully return to the $M_S = 1$ states at large times.  However, over the time scale of microseconds, effects of decoherence, which were not taken into account in this calculation, may be more significant than the small changes observed in this calculation (see Sec.~\ref{sec:decoherence} below).

The adiabaticity can be turned off for any combination of strain and electric fields. 
The transverse magnetic field $(B_x,B_y)$ needed to eliminate the adiabaticity in the general case where $\epsilon=\epsilon_1+i\epsilon_2$ is complex must satisfy the equations $B_y^2-B_x^2=\epsilon_1/\eta$ and $2B_xB_y=-\epsilon_2/\eta$. For example, if $\epsilon=i\epsilon_2$ is purely complex, the adiabaticity is eliminated by $B_x= -B_y= \pm \epsilon_2/2\eta$.
Once the effective coupling $\epsilon({\bf B}_{\perp})$ is eliminated by controlling the transverse magnetic field,  the adiabaticity is turned off for {\em any} radio-frequency $\omega$ and {\em any} magnetic field strength $B_0$.

\section{Dynamics including hyperfine interaction} \label{sec:hf}
\label{sec:hyperfine}

In this section we study the effect of the hyperfine interaction on the adiabaticity of NV spin dynamics. In particular, we calculate the dynamics of the spin including the hyperfine structure when a transverse magnetic field is used for cancelling the effect of $\epsilon_{es}$ and eliminating the gap between the adiabatic energy eigenvalues $E_{\pm}({\bf B})$. 
 
If the hyperfine interaction with the ${}^{14}$N nuclear spin ($I=1$) is included, the Hamiltonian can be written as a 9$\times$9 matrix,
\begin{eqnarray}  \label{Eq:H_nv_nucl}
    H_{\mathrm{hf}}(t)&=& H(t) \otimes {\bf 1}^{\mathrm{nucl}}_{3\times3}  + {\bf 1}_{3\times3} \otimes H^{\mathrm{nucl}}(t) + H^{\mathrm{int}},
\end{eqnarray}
where
\begin{eqnarray}  \label{Eq:H_nucl}
    && H^{\mathrm{nucl}}(t) = {\cal D}_n (I_z^2 -\frac{2}{3} {\bf 1})  + \mu_n \, {\bf I} \cdot {\bf B}(t) ,
\end{eqnarray}
and
\begin{eqnarray}  \label{Eq:H_int}
    &&H^{\mathrm{int}} = {\cal A}_\parallel S_z \otimes I_z + {\cal A}_\perp (S_x \otimes I_x + S_y \otimes I_y).
\end{eqnarray}
where  the zero-field splitting of the nuclear spin is ${\cal D}_n = h\times 5.0$ MHz, the nitrogen magnetic moment is $\mu_n = (0.403/1837) \mu_B = 0.612$ kHz/G, and the hyperfine splitting factors are ${\cal A}_\parallel = h\times 2.2$ MHz and ${\cal A}_\perp = h\times 2.1$ MHz.  
 
In the same way as for the interaction of the electronic spin with the magnetic field, the transverse magnetic field has a second-order effect on the nuclear spin, which is smaller by a factor of $\mu_n/2{\cal D}_n$ than the effect of the longitudinal field. Moreover, this factor is an order of magnitude smaller than the equivalent factor $\mu/2{\cal D}$ of the electronic spin, so that the overall effect is is about 4 orders of magnitude smaller than that of the electronic spin. In addition, the effect of the transverse hyperfine interaction couples in its second order the states $|M_S=1,M_I=-1\rangle$ and $|M_S=-1,M_I=1\rangle$ with a coupling strength of ${\cal A}_{\perp}^2/2{\cal D} \sim 1$\,kHz, which is smaller than any other interaction by 3 orders of magnitude.  We therefore neglect here the effect of the transverse interactions. 

In this approximation the Hamiltonian can be decoupled into three blocks of 2$\times$2 matrices for $M_S=\pm 1$ levels and three blocks of 1$\times$1 for $M_S=0$ levels. The 2$\times$2 matrices can be written as
\begin{eqnarray} \label{eq:H_MI}
H_{\pm 1,M_I} &&= \left[\frac13{\cal D}+{\cal D}_n(M_I^2-\frac23) +\mu_nM_I B_z \right] {\bf 1}_{2\times2} + \nonumber \\
&&+\left(\!\!\begin{array}{cc} {\cal A}_{\parallel}M_I+\mu B_z& \epsilon({\bf B}_{\perp}) ^*\\ \epsilon({\bf B}_{\perp}) & -{\cal A}_{\parallel}M_I-\mu B_z \end{array} \!\!\right). 
\end{eqnarray}
The adiabatic eigenstates and eigenvalues of this Hamiltonian are then exactly the same as those of the two-level Hamiltonian [Eq.~(\ref{Eq:H_nv_Bz}), except that the position of the pseudo-crossing as a function of $B_z$ is shifted by $ \Delta B_z(M_I)=-{\cal A}_{\parallel}/\mu$ and the energy eigenvalues of each $M_I$ pair of levels is shifted by ${\cal D}_n(M_I^2-\frac23)$.  This implies that when the magnetic field is swept through the pseudo-crossing the transitions between the $M_s=\pm 1$ states occur at different times for each $M_I$ state. 
Transverse field components $B_x,B_y$ couple between different $M_I$ blocks, but their effect is expected to be negligible when $|B_x|,|B_y|\ll {\cal D}_n/\mu_n$, i.e., when these components are less than hundreds of Gauss.

\begin{figure}%[ht]
\centering
\includegraphics[width=0.8\linewidth]{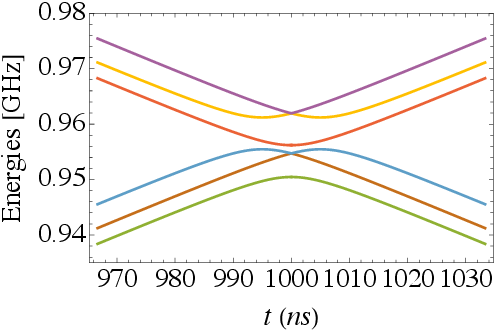}
\caption{Adiabatic energy eigenvalues for the case of a magnetic field ${\bf B}(t)=\hat{\bf z}B_0\cos\omega t$ with all parameters as used in Figs.~\ref{Fig_1}, for the 9-level (practically, 6-level) Hamiltonian in Eq.~(\ref{Eq:H_nv_nucl}). The three eigenvalues associated with the $M_S = 0$ electronic state  whose energies are near -1.913 GHz lie outside the plot range.  The 6 energy eigenvalues consist of three pairs of eigenstates of the Hamiltonians $H_{\pm 1,M_I}$ [Eq.~(\ref{eq:H_MI})] for $M_I=0,\pm 1$. The two energy eigenvalues of each pair are similar to those of the pair of electronic levels shown in Fig.~\ref{Fig_1}, except for an energy shift ${\cal D}_n(M_I^2-\frac23)$ and a shift along the $t$ axis. As discussed in the text, terms in the Hamiltonian that involve transitions between pairs of levels with different $M_I$ are negligible and the pseudo-crossing within adiabatic levels of the same pair is similar to the one described by the electronic levels only.  
}
\label{Fig_1hyp}
\end{figure}

Figure~\ref{Fig_1hyp} shows the adiabatic eigenstates of the 9-level Hamiltonian as calculated for the same parameters and on the same scale used in Fig.~\ref{Fig_1}.  The three levels that come from $M = 0$ state are not shown in the figure because they are far removed from the 6 levels shown and effectively do not couple to the 6 levels shown. The 6 curves appearing in this figure have almost exactly the same shape as those of the two curves corresponding to the electronic states, except that the curves of the hyperfine states are shifted in energy and in time due to the change of the offset magnetic field. 

\begin{figure}%[hb]
\centering
\subfigure[]
{\includegraphics[width=0.5\linewidth]{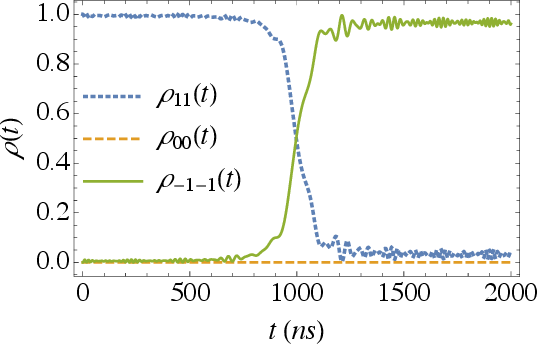}}
\centering
\subfigure[]
{\includegraphics[width=0.5\linewidth]{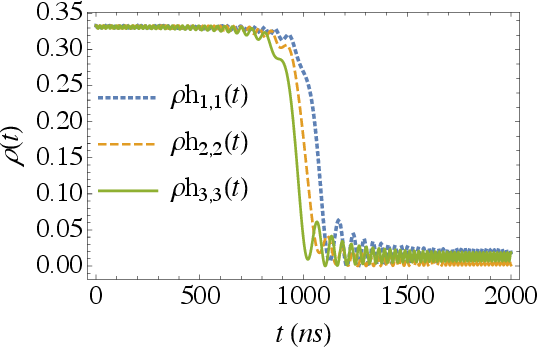}}
\subfigure[]
{\includegraphics[width=0.5\linewidth]{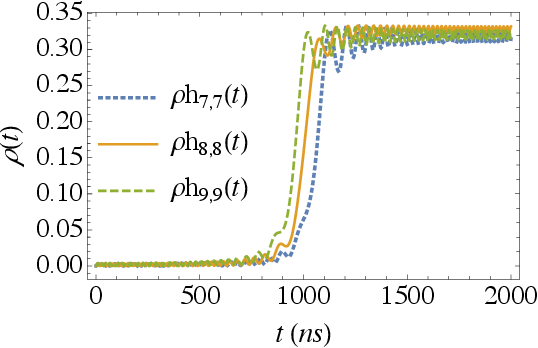}}
\caption{Almost fully adiabatic dynamics for the same parameters  used in Fig.~\ref{Fig_1hyp} (including the nuclear degrees of freedom, compare to Fig.~\ref{Fig_2} in Appendix~\ref{app:longi}) with the population initially distributed equally in the hyperfine sublevels of  the $M_S = 1$ level. (a) sum over the hyperfine levels. Here $\rho_{j \, j}$, $j = +1, 0, -1$, is a sum over all three $M_I$ sublevels of the $M_S = j$ manifold. (b) $\rho_{k,k}(t)$ for $k = 1, 2, 3$ --  the probabilities for the hyperfine levels with $M_S = 1$. Here the subscripts $k$ refer to the 9 levels $k=5-3M_S-M_I$ ($k = 1, \ldots, 9$).  These curves are labeled $\rho h_{k,k}(t)$ in the figure labels.  (c) Same as (b) except here, only the probabilities $\rho_{k,k}(t)$ for $k = 7, 8, 9$ associated with the $M_S = -1$ level are shown.  Probabilities associated with the $M_S = 0$ level are zero throughout the dynamics and are not shown.
For the sum probabilities in (a) the transition from $M_S=1$ to $M_S=-1$ is a bit less sharp than in Fig.~\ref{Fig_2} of Appendix~\ref{app:longi}  and the oscillations are smeared due to the offset in the time of transition among the three hyperfine levels, as shown in (b) and (c) and as explained in Fg.~\ref{Fig_1hyp}. 
}
\label{Fig_5}
\end{figure}

Now we examine the dynamics of the system with the same parameters used in the previous section, but  including the nuclear degrees of freedom, i.e., we treat the system with hyperfine interactions included.  Figure \ref{Fig_5} shows the 9-level population dynamics corresponding to the parameters used in Fig.~\ref{Fig_1hyp}, which give rise to an adiabatic  dynamics where all the population is transferred from the 3 hyperfine levels of the electronic level $M_S=1$ to the corresponding levels of $M_S=-1$ when the longitudinal magnetic field changes sign (no transverse fields are applied). 
The curves in Fig.~\ref{Fig_5}(a)  show the populations versus time summed over the hyperfine levels.  
When comparing to the two-level dynamics where the nuclear spin degrees of freedom are neglected  (see Fig.~\ref{Fig_2} in \ref{app:longi}), we see that  the curves are very similar, except that in the presence of nuclear spin the oscillations are more ragged and the slopes of the curves at the transition point are a bit smaller because the populations in the individual hyperfine pairs of  levels [see Figs.~\ref{Fig_5}(b) and (c)] evolve with slight temporal offsets from one another, hence the features of the curve with a sum over hyperfine levels are somewhat smeared.
 
\begin{figure}% [hb]
\centering
\subfigure[]
{\includegraphics[width=0.6\linewidth]{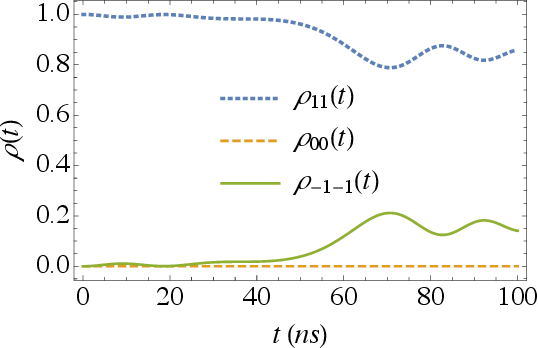}}
\centering
\subfigure[]
{\includegraphics[width=0.6\linewidth]{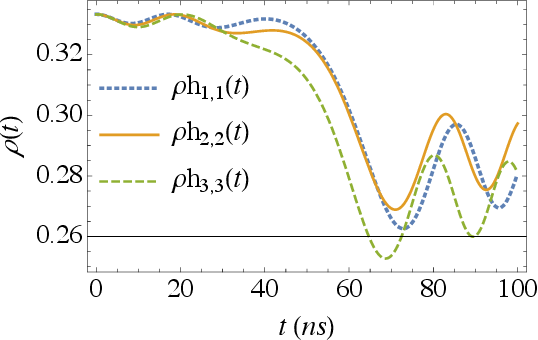}}
\subfigure[]
{\includegraphics[width=0.6\linewidth]{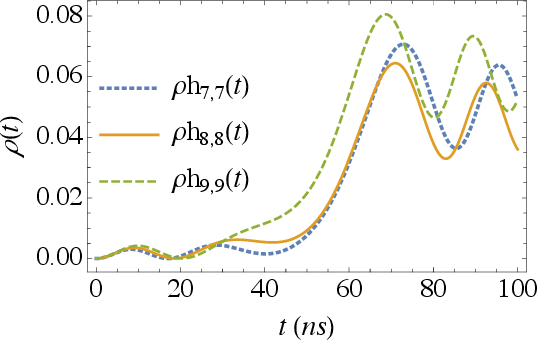}}
\caption{Non-adiabatic dynamics obtained with an angular frequency $\omega$ of the oscillating magnetic field 20 times larger than that used in Figs.~\ref{Fig_1} and Figs.~\ref{Fig_1hyp},\ref{Fig_5}, corresponding to an adiabaticity parameter $\beta=5.41\gg 1$. All the other parameters are identical. (To compare with Fig.~\ref{Fig_5}, scale the abscissa by a factor of 20).
 (a) Sum over the hyperfine levels.
 (b) $\rho_{k,k}(t)$ for $k = 1,2, 3$ associated with the $M_S = +1$ manifold.  (c) $\rho_{k,k}(t)$ for $k = 7, 8, 9$ associated with the $M_S = -1$ manifold.  
}
\label{Fig_6}
\end{figure}

Figure \ref{Fig_6} shows the 9-level population dynamics when the angular frequency of the oscillating magnetic field is increased to 20 times its value in Fig.~\ref{Fig_5}, giving rise to non-adiabatic dynamics ($\beta\gg 1$). 
As in Fig.~\ref{Fig_5}, Fig.~\ref{Fig_6}(a) shows the sum populations corresponding to the electronic states $M_S=\pm 1$ and Figs.~\ref{Fig_6}(b) and (c) show the populations of the different hyperfine levels. Comparison to the two-level dynamics in Fig.~\ref{Fig_3} in \ref{app:longi} reveal the same features discussed above with respect to Fig.~\ref{Fig_5}. 

\begin{figure}% [ht]
\centering
\subfigure[]
{\includegraphics[width=0.5\linewidth]{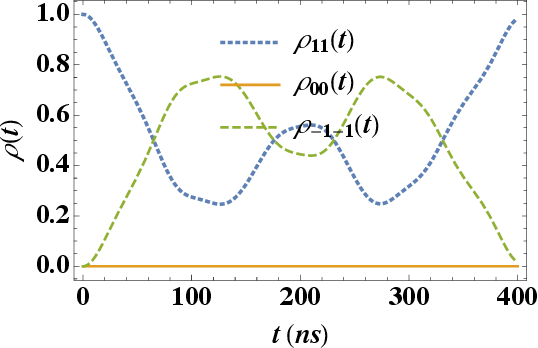}}
\centering
\subfigure[]
{\includegraphics[width=0.5\linewidth]{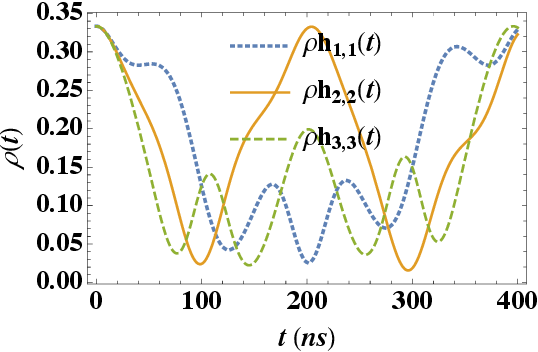}}
\subfigure[]
{\includegraphics[width=0.5\linewidth]{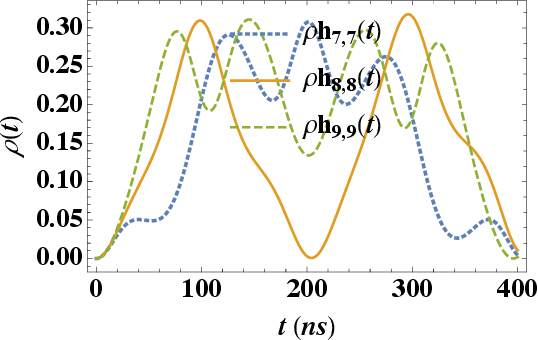}}
\caption{Non-adiabatic dynamics obtained with the same parameters used in Fig.~\ref{Fig_6}, but the magnetic field strength is reduced by a factor of 5, i.e., $\mu B_z = 2 \pi \times 4.8$ MHz (corresponding to a magnetic field of 2 Gauss). The adiabaticity parameter is $\beta=1.08\sim 1$, the slope of the eigenenergies versus time is slower than in Fig.~\ref{Fig_6} and the dynamics here are less non-adiabatic. 
Here the time of the dynamical calculation has been extended by a factor of four (the dynamics are followed for four periods of the radio-frequency field).
(a) The sum over the hyperfine levels, with the population initially distributed equally among the hyperfine $M_I$ sub-levels of  the $M_S = 1$ level. (b) $\rho_{k,k}(t)$ for $k = 1,2, 3$ associated with the $M_S = +1$ manifold.  (c) $\rho_{k,k}(t)$ for $k = 7, 8, 9$ associated with the $M_S = -1$ manifold.  Probabilities associated with the $M_S = 0$ level are almost zero throughout the dynamics.
}
\label{Fig_7}
\end{figure}
 
Figure \ref{Fig_7} shows the 9-level population dynamics in the intermediate case where the adiabaticity parameter is close to unity, $\beta\approx 1$, corresponding to the angular frequency  $\omega$ as in Fig.~\ref{Fig_6} and the magnetic field amplitude $B_0$ decreased by a factor of 5 to $B_0=2$\,G. 
Note that we have extended the range of time in the dynamical calculation to a final time of $t = 400$ ns (recall that the first pseudo-crossing occurs at 50 ns for this set of parameters).  Here too, the curves of the different hyperfine levels appear at a time offset from each other and hence the sum curves show the same smearing features shown in Figs.~\ref{Fig_5} and~\ref{Fig_6}.  Except for these features, the sum population behaves quite similarly to that calculated for the model that neglects the hyperfine interaction.

\begin{figure} %[hb]
\centering
\subfigure[]
{\includegraphics[width=0.5\linewidth]{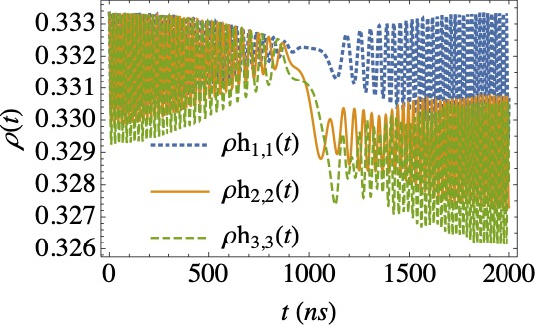}}
\centering
\subfigure[]
{\includegraphics[width=0.5\linewidth]{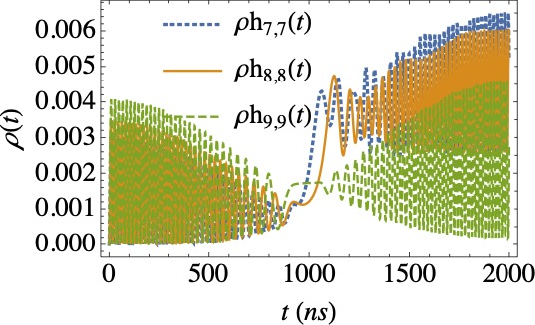}}
\centering
\subfigure[]
{\includegraphics[width=0.5\linewidth]{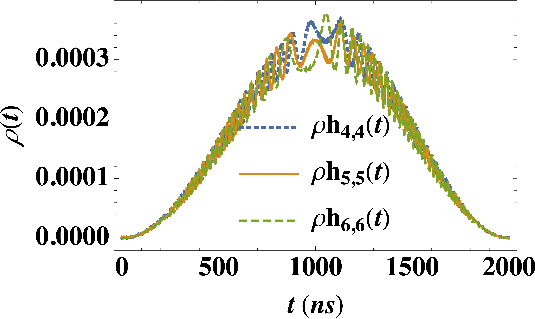}}
\caption{9-level non-adiabatic dynamics obtained with the same parameters used in Fig.~\ref{Fig_non-adiab-pop} but a transverse magnetic field exactly tuned to the specific field $B_c$ that eliminates the energy gap between the adiabatic eigenvalues. (a) $\rho_{k,k}(t)$ for $k = 1,2, 3$ associated with the $M_S = +1$ manifold.  (b) $\rho_{k,k}(t)$ for $k = 7, 8, 9$ associated with the $M_S = -1$ manifold.  (c) $\rho_{k,k}(t)$ for $k = 4, 5, 6$ associated with the $M_S = 0$ manifold.
Transitions to the $M_S=-1$ manifold are of the order of 1\%, showing that adiabaticity cancellation by a transverse field is effective even if hyperfine interactions are taken into account. 
}
\label{Fig_non-adiab-pop-9-level}
\end{figure}

Finally, in Fig.~\ref{Fig_non-adiab-pop-9-level} we present the dynamics of the spin including the hyperfine structure when the transverse magnetic field is set to cancel the effect of $\epsilon_{es}$ and eliminate the gap between the adiabatic energy eigenvalues. It is shown that the dynamics becomes completely non-adiabatic, as found for the three-level electronic system without considering hyperfine interactions. However, in some applications based on the use of transverse fields to cancel the adiabaticity, it might be necessary to take the hyperfine structure into account if high precision is needed. 

\section{Dynamics including decoherence} \label{sec:decoherence}

Open systems, i.e., systems that interact with their environment, undergo dephasing, decoherence and relaxaton.  For systems that are coupled to Gaussian white noise,  the stochastic dynamics can be described using the Schr\"{o}dinger--Langevin equation \cite{vanKampenBook}.  If one averages over the stochasticity, one can obtain a Markovian quantum master equation for the density matrix $\rho(t)$ with Lindblad operators ${\cal V}_j$ \cite{vanKampenBook, master_eq}:
\begin{eqnarray}  \label{Eq:master}
    {\dot \rho} &=& -i [H(t), \rho(t)] \\ \nonumber 
    &+& \frac{1}{2} \sum_j \xi_{0,j}^2
    \left(2{\cal V}_j \rho(t) {\cal V}^{\dag}_j - \rho(t) {\cal
    V}^{\dag}_j {\cal V}_j - {\cal V}^{\dag}_j {\cal V}_j \rho(t)
    \right) .
\end{eqnarray}
The Lindblad coefficients (volatilities) $\xi_{0,j}$ specify the strength of the white noise. We may assume that the interaction of the three-level system of the electronic ground state of the NV center with the environment is equivalent to the interaction of the NV spin with magnetic noise arising from a bath (i.e., from an environment) or several baths. The Lindblad operators for this case can be taken to be the three spin-1 operators, ${\cal V}_j = S_j$ ($S_x, S_y, S_z$ for $j =x, y, z$).  A NV near the diamond surface experiences a bath due to noise originating from the diamond surface that has fast correlation times, perhaps even as fast as $\tau_c \approx 10^{-11}$ s ~\cite{Rosskopf14}.  Therefore, shallow NVs in diamond have an environment correlation time comparable to, or shorter than, the energy splitting parameter ${\cal D}^{-1}$, and one approaches the white noise limit.  
Hence, shallow NVs can be modeled by the Lindblad master equation (\ref{Eq:master}) \cite{Ajisaka_16}. Note that experimental NV coherence times reported in Ref.~\cite{Rosskopf14} are on the order of tens to hundreds of $\mu$s.
 
For simplicity we take the interaction of the NV with the environment to be isotropic, such that  the volatilities $\xi_{0,j}$ are equal for $j=x, y, z$. In this case the Lindblad part of the master equations is given by the following decoherence and relaxation terms,
\begin{eqnarray}
\dot{\rho}_{11}^{\gamma} &=& \gamma(\rho_{00}-\rho_{11}) \nonumber \\
\dot{\rho}_{00}^{\gamma} &=& \gamma(\rho_{11}+\rho_{-1\,-1}-2\rho_{00}) \nonumber \\
\dot{\rho}_{-1\,-1}^{\gamma} &=& \gamma(\rho_{00}-\rho_{-1\,-1}) \nonumber \\
\dot{\rho}_{0\,\pm 1} &=& \gamma(\rho_{\mp 1\,0}-2\rho_{0\,\pm 1}) \nonumber \\
\dot{\rho}_{1\,-1}^{\gamma} &=& -3\gamma\rho_{1\,-1}
\end{eqnarray}
where $\gamma=2\xi_{0,j}^2$. These equations (without the LZ dynamics) can be solved analytically.  The dynamics of the diagonal elements of the density matrix (the occupation probabilities of the levels) leads to their relaxation into a state of equal population, $\rho_{i i}=1/3$. This evolution of the populations involves terms that decay exponentially like $e^{-\gamma t}, e^{-2\gamma t}$ and $e^{-3\gamma t}$. The decoherence of the off-diagonal elements is uncoupled from the relaxation of the diagonal elements. The LZ dynamics may create superpositions of the $M_S=\pm 1$ levels such that $\rho_{1\,-1}\neq 0$. This coherence decays at a rate of $3\gamma$, such that the coherence time of the superposition is shorter than the relaxation time. 
 
\begin{figure} %[ht]
\centering
\includegraphics[width=0.8\linewidth]{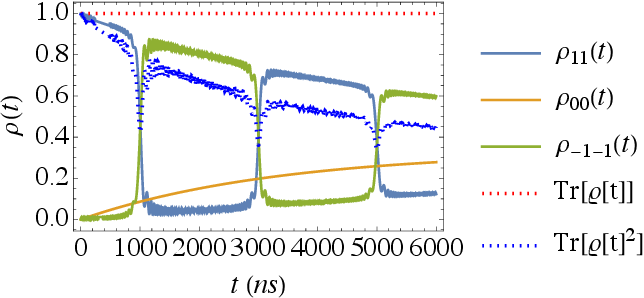}
\caption{LZ dynamics in the presence of decoherence. Probabilities $\rho_{11}$ (pale blue curve on top left), $\rho_{00}$ (orange curve on bottom left) and $\rho_{-1 \, -1}$ (green curve on bottom right) versus time [same parameters as in the unitary case (without decoherence) shown in Fig.~\ref{Fig_1}.  We use isotropic noise with volatilities $\xi_{0,j}^2 = 1 \times 10^5$ s$^{-1}$, $j =x, y, z$, which corresponds to $\gamma = 2 \times 10^5$ s$^{-1}$.  The red curve shows the sum of the probabilities versus time, ${\mathrm{Tr}}[\rho(t)]$, which remains unity, and the blue curve shows the purity, $p(t) \equiv {\mathrm{Tr}}[\rho^2(t)]$.}
\label{Fig_decoherence}
\end{figure}

To demonstrate the LZ dynamics in the presence of relaxation and decoherence we take $\gamma = 2 \times 10^5$ s$^{-1}$, corresponding to a $\xi_{0,j}^2 = 1 \times 10^5$ s$^{-1}$ for $j = x, y, z$ and a relaxation time of $5 \times 10^{-6}$ s.  Hence relaxation and decoherence have a significant effect on the time scale of a few periods of oscillation of the magnetic field. In Fig.~\ref{Fig_decoherence} we show the occupation probabilities $\rho_{11}(t)$, $\rho_{00}(t)$ and $\rho_{-1 \, -1}(t)$ versus time.  
Relaxation is apparent in each of the probabilities and can be compared with the results in Fig.~\ref{Fig_2} of Appendix~\ref{app:longi} obtained for the same system parameters but without the presence of noise.  At very long time, the population is equally distributed among all three levels (the environment is not a thermal bath).  Of course the total probability $\sum_{i} \rho_{ii}(t)$ remains unity (see red curve in Fig.~\ref{Fig_decoherence}) throughout the course of the dynamics. The purity $p(t) \equiv {\mathrm{Tr}}[\rho^2(t)]$ decreases with time, but not monotonically. Asymptotically at large time, the purity reaches a value of 1/3, and this despite the fact that the $M_S = 0$ level would not participate in the dynamics at all and would remain with zero population where it not for the decoherence.  (Note however that the populations will be given by the Boltzmann distribution at very long time.) Figure~\ref{Fig_decoherence} shows that after a time comparable to the coherence time of $t = 10 \, \mu$s each of the populations of the three levels is already close to 1/3 and hence the effect of LZ transitions on the populations is damped. 
 
Quite generally, for Lindblad master equations describing three-level systems, the decoherence cannot be modeled using a single exponential decay rate $\gamma$.  A more complicated temporal dependence results because of the presence of multiple decay timescales.   The decoherence behavior can be understood as follows. For a time-independent Hamiltonian, each of the matrix elements of the density matrix can be expressed as
$$
\rho_{\alpha\beta}(t) =a_{0,\alpha\beta} +\sum_{i=1}^8 a_{i,\alpha\beta} \exp (-\gamma_i t) ,\ \ 
(\alpha,\beta=1,2,3)
$$
where the $\gamma_i$ ($i = 0, 1, \ldots, 8$) are the 9 eigenvalues of the 9$\times$9 Liouvillian operator,  and $\gamma_0$ is zero.  The real parts of $\gamma_i$ determine decay rates and the imaginary parts determine energy eigenvalue differences, $a_i$ ($i = 1, \ldots, 8$) are the amplitude coefficients, and the coefficient $a_0$ corresponds to the amplitude of the steady state whose existence is guaranteed by trace preservation.  Hence, for a time-independent three-level system, the maximum number of possible timescales that determine the population decay and the coherence dynamics is 8 (the number of non-zero eigenvalues), but there may be a lower the number due to symmetry. For a time-dependent Hamiltonian in the adiabatic regime, the eigenvalues $\gamma_i$ and the amplitudes $a_i$ are time-dependent, but an adiabatic expansion can still be carried through \cite{Band_92}.  In any case, it is clear from this analysis that more than one decay rate is in general required to describe the decoherence of a three-level system.  Our use of equal decay rates for three Lindblad operators while setting the remaining decay rates to zero is therefore an approximation.

\section{Summary and Conclusions} \label{sec:S&C}
 
We studied the dynamics of negatively charged NV color centers in diamond under the influence of a time-dependent magnetic field which induces energy eigenvalue pseudo-crossings.  The avoided crossing is due to coupling between spin states $|M_S=\pm1\rangle$ caused by strain and local electric fields at the site of the NV center, but also depends on the components of the magnetic field vector in the plane perpendicular to the NV axis. As long as these magnetic fields are stationary or slowly varying (in the radio-frequency range) the three-level Hamiltonian of the triplet ground state of the NV center may be reduced into an effective two-level Hamiltonian for the magnetic field sensitive states $|M_S=\pm1\rangle$ while the state $|M_S=0\rangle$ is eliminated from this spin dynamics {\c (but the effects of the $|M_S=0\rangle$ {\em are} incorporated into the effective two-level Hamiltonian).  A transverse static magnetic field can therefore tune the effective adiabaticity of the LZ dynamics and make the effective coupling between states vanish.  In this case the avoided crossing between the spin states is completely eliminated because the effective off-diagonal coupling can be made to vanish, and the dynamics becomes non-adiabatic even if the longitudinal field is swept through zero very slowly.  It is important to note that once the avoided crossing is eliminated by controlling the transverse magnetic field, the adiabaticity is turned off for {\em any} radio-frequency $\omega$ and {\em any} magnetic field strength $B_0$.

By tuning the transverse magnetic field, one can probe the coupling between the $M_S = \pm 1$ states and extract information about the direction and strength of the internal strain and electric fields that cause the zero-field splitting between the otherwise degenerate states $M_S=\pm 1$, i.e., transverse magnetic field tuning may be useful for sensing (measuring) stress and electric fields.

We also studied the effect of the hyperfine structure due to the nitrogen $^{14}$N nuclear spin ($I=1$) on the spin dynamics and found that the Hamiltonian for these 6 levels corresponding to the $M_S=\pm 1$ states can be decoupled into three pairs of levels with the same nuclear spin projection $M_I$, as the nuclear spin does not change its state when the magnetic field is swept through zero. Each of these pairs has exactly the same structure as the pair of levels studied without the nuclear spin, except for a magnetic field offset induced by the hyperfine interaction. It follows that for each of the pairs transitions between the $M_S = \pm 1$ states, or equivalently Landau-Zener transitions between the adiabatic levels, occur at slightly different times for different nuclear spin states when the external oscillating field is swept through the pseudo-crossings of the pairs of levels. When all nuclear spin states are occupied, it therefore appears that the transitions are smeared. 

Moreover, we studied the effects of coupling of shallow NVs near diamond surfaces to an environment modeled by a spin-bath with short fluctuation times (Gaussian white noise). Such coupling leads to decoherence and relaxation of the spin population into a mixed state with equal occupation of all $M_S$ states. We showed that the combination of the relaxation and decoherence (with or without the effect of the oscillating field) gives rise to a non-monotonic decrease of the purity of the state and the purity goes to 1/3 at very long times, i.e., all the $M_S$ spin components eventually become equally populated (at very large times and at finite temperature, the populations are given by the Boltzmann distribution).

Beyond the prospects of measuring local or external stress and electric fields by using the adiabaticity properties of the spin dynamics, as mentioned above, this work may yield additional applications. Controlling the coupling between the $|M_S=\pm1\rangle$ electronic levels is particularly important in magnetometry with NV centers in diamond, where this coupling suppresses the linear response of the Zeeman energies to weak magnetic fields. This can be dealt with by adding a strong magnetic bias field or by using specially manufactured diamonds with a reduced strain \cite{Zheng_19}. Our proposal to use a perpendicular magnetic field to eliminate the coupling by a transverse magnetic field may serve as an alternative for achieving zero-field sensitivity with an almost purely linear response for a given orientation of the NVs in diamond. 

More generally, understanding and controlling the adiabaticity of spin dynamics may assist in designing specific schemes for quantum control of spin states, especially when they are coupled to other degrees of freedom of the system.  In particular, this may be crucial to quantum devices based on levitated solid nano-objects when the spin states together with magnetic fields are used for controlling their external degrees of freedom~\cite{Hsu_2016,Delord_2020, Margalit2021}. 
These objects rotate in space so that the axis direction of the spin (NV) centers change orientation, hence, in the frame of reference of the spin center, the longitudinal and perpendicular magnetic field components vary in time even if the magnetic field is stationary in the lab frame. The adiabaticity of spin dynamics under these conditions may be crucial in future developments in cooling, trapping and manipulating these objects.

\bigskip
 
%\begin{acknowledgments}
We thank Yosef Rosenzweig, Yechezkel Schlussel and Ron Folman for initiating the motivation for this paper and for useful conversations.  This work was supported in part by grants from the DFG through the DIP program (FO703/2-1).
%\end{acknowledgments}
 
\appendix
\section{Definition of adiabaticity}
\label{app:adiabaticity}

When considering the dynamics of a system based upon the time-dependent Schr\"{o}dinger equation, $i\hbar \, d\psi/dt = H(t)\psi$, with a time-dependent Hamiltonian $H(t)$ that varies ``slowly'', the adiabatic theorem \cite{Band_Avishai} tells us that  If the system is initially in an eigenstate $u_n(t_0)$ of the initial Hamiltonian, $H(t_0)$, it remains in the eigenstate $u_n(t)$ of the instantaneous Hamiltonian $H(t)$ ($t>t_0$) that at each instant of time $t$ satisfies
\begin{equation} \label{Eq:Adiab.0}
    H(t) u_n(t) = E_n(t) u_n(t) ~,
\end{equation}
and is therefore called ``an adiabatic state". If the initial state is a superposition of initial eigenstates, $\psi_{\mathrm{in}} =\psi(t_0) = \sum_n c_n(t_0) u_n(t_0)$, it evolves to a superposition of the instantaneous eigenstates
$\psi(t) = \sum_n c_n(t_0) u_n(t) \exp [\frac{-i}{\hbar} \int_{t_0}^{t} dt' \, E_n(t') ]$ with the same probabilities (adiabatic evolution) if the Hamiltonian varies sufficiently slowly, in the sense that \cite{Band_Avishai_ibid}
\begin{equation} \label{Eq:Adiab.1}
   \int_{t_0}^t dt' \, \left| \frac{\langle u_i(t')| \frac{dH(t')}
   {dt'} |u_n(t')\rangle} {[E_n(t')-E_i(t')]} \right| \ll 1 \, \, \rm{ for } \, \, n \ne i. %(9.43)
\end{equation}
The simplest model for LZ dynamics is a two-level system with Hamiltonian $H(t) = \left( \! \!
\begin{array}{cc}
\varepsilon_1 & V^*  \\
V & \varepsilon_2 + \alpha t
\end{array} \! \! \right)$. The higher energy eigenstate at $t\to -\infty$,  $|a(-\infty)\rangle= (1,0)^{\dag}\equiv |1\rangle$ with $E_a(t\to -\infty)= \varepsilon_1$ evolves continuously with an infinitesimally small rate of change, $\alpha \to 0$, into the higher energy eigenstate $|a(t\to +\infty)\rangle = (0,1)^{\dag}\equiv |2\rangle$ with energy $E_a(t\to +\infty)=\varepsilon_2+\alpha t$ (see the green curve in Fig.~\ref{Fig_1}).  The Landau-Zener transition probability $P_{LZ}$,
\begin{equation}   \label{Eq:LZ}
    P_{LZ} = \exp\left( \!  -2\pi \frac{ V^2}{ \hbar \alpha} \! \right) ~.
\end{equation}
of finding the system in the other {\em adiabatic} state, $|b\rangle$, which is the lower energy state that is orthogonal to $|a\rangle$ and coincides with the {\rm diabatic} state $|1\rangle$ at $t\to \infty$ (see the yellow curve in Fig.~\ref{Fig_1}), is then negligible because the adiabatic theorem ensures that the system stays on the initial {\em adiabatic} state.  For finite $\alpha$, the transition probability $P_{LZ}$ at the {\em final time} depends exponentially on the inverse rate of change of the energy difference. The adiabatic transition probability $P_{ad}$ of finding the system in the {\em adiabatic} state $|a\rangle$, which coincides with the {\em diabatic} state $|2\rangle$ at $t\to \infty$, is
\begin{equation}   \label{Eq:ad}
    P_{ad} = 1-P_{LZ} = 1 - \exp\left( \!  -2\pi \frac{ V^2}{ \hbar \alpha} \! \right) ~.
\end{equation}
$P_{ad}$ is almost unity for small $\alpha$, as the adiabatic theorem ensures that the system stays on the initial {\em adiabatic} state, which amounts, in this example, to flipping the {\rm diabatic} state.

\section{Derivation of the effective $2\times 2$ Hamiltonian}
\label{app:Heff}

We consider a Hilbert space that contains two parts: one part that we are particularly interested in, and another part whose energy is far removed from the first.  We define a projection operator $P$ into the first part and the complementary projection operator $Q=\hat{1}-P$ that projects onto the second part.  The Schr\"odinger equation $i\hbar\partial_t\psi=H\psi$ can then be separated as follows:
\begin{eqnarray} 
i\hbar\partial_t P\psi=PHP\cdot P\psi+PHQ\cdot Q\psi, \label{eq:dtPpsi} \\
i\hbar\partial_t Q\psi=QHQ\cdot Q\psi+QHP\cdot P\psi.
\label{eq:dtQpsi}
\end{eqnarray}
If the Hamiltonian $H$ is stationary and we seek to find eigenstates and eigen-energies of the system, then we replace $i\partial_t\psi\to E\psi$. We then obtain the following from Eq.~(\ref{eq:dtQpsi}):
\begin{equation} Q\psi=\frac{1}{E-QHQ}QHP\cdot P\psi. \label{eq:Qpsi} 
\end{equation}
By substituting Eq.~(\ref{eq:Qpsi}) into Eq.~(\ref{eq:dtPpsi}) we obtain
\begin{equation} E P\psi=\left(PHP+PHQ\frac{1}{E-QHQ}QHP\right)P\psi. 
\label{eq:EPpsi} \end{equation}
This equation is exact, and has a practical advantage for approximating the eigen-energies of the first part of Hilbert space when $QHQ$ is far larger in magnitude than all the other parts of the Hamiltonian. In this case there are solutions for which $|E|\ll |QHQ|$, corresponding to eigenstates with small occupation of $Q\psi$. In this case a good approximation for the energy eigenstates and eigenvalues is obtained by setting $E\to 0$ in the right-hand-side of Eq.~(\ref{eq:EPpsi}). Moreover, an improved approximation is obtained by substituting this zero-order approximation for $E$ into the right hand side of the equation. 

In the time-dependent case, where the rate of change of $PHP$ and $QHP$ is much smaller than the $QHQ/\hbar$, we use a lowest order approximation that ignores the time-dependence of $Q\psi$ on the left-hand-side of Eq.~(\ref{eq:dtQpsi}) and obtain and effective Hamiltonian 
\begin{equation} H_{\rm eff}=PHP-PHQ\frac{1}{QHQ}QHP. 
\end{equation}
This Hamiltonian is equivalent to the zeroth order approximation in the stationary case, Eq.~(\ref{eq:EPpsi}), where $E\to 0$ in the right-hand-side.

\section{Dynamics with longitudinal magnetic field only}
\label{app:longi}

For demonstrating the basic principles of LZ dynamics in our system we consider  a simple situation where the magnetic field is parallel to the axis of the NV center ${\bf B}=B_z\hat{z}$. In this case the $M_S=0$ level is not coupled to the $M_S=\pm1$ levels and the 3$\times$3 Hamiltonian matrix for the levels of interest $M_S=\pm1$ reduces to a 2$\times$2 matrix with two adiabatic eigenstates whose minimum energy difference is given by the zero field splitting $|\epsilon|$. This implies that the dynamics upon slowly sweeping the magnetic field through ${\bf B}=0$ is adiabatic and diabatic Landau-Zener transitions are suppressed.  When the external magnetic field ${\bf B}(t)$ is in the $\hat{z}$ direction, i.e., along the axis of the NV, the energies of the levels $M_S=\pm 1$ depend linearly on the magnetic field component $B_z$ while the transverse components $B_x,B_y$ couple between these sub-levels and the level $M_S=0$, whose energy is far by ${\cal D}$. As we see below the relevant energy scale for LZ dynamics is governed by $\epsilon$, which is 3 orders of magnitude smaller than the splitting ${\cal D}$. The effect of transverse fields of the same order as $\epsilon$ on the dynamics is then smaller by a factor of $\epsilon/{\cal D}\sim 10^{-3}$ than the effect of other energy scales and therefore at this stage we consider only a magnetic field in the $\hat{z}$ direction, while the effects of transverse fields is discussed in Sec.~\ref{sec:tune_adiab}. 
If we consider only the axial magnetic field $B_z$ then the dynamics of the levels $M_S=\pm1$ is determined by a 2$\times$2 Hamiltonian 
\begin{equation} \label{Eq:H_nv_Bz}
H(t)=\left(\begin{array}{cc} \mu B_z(t) & \epsilon^* \\ \epsilon & -\mu B_z(t)\end{array}\right)
+\frac13{\cal D}  {\bf 1}. 
\end{equation}
For this Hamiltonian, with trigonometric magnetic field time dependence $B_z(t)= B_0 \cos(\omega t)$, the off-diagonal coupling potential of the LZ problem [Eq.~(\ref{Eq:LZ})] is given by $V = \epsilon$ and the rate of energy change is $\alpha = 2\mu B_0 \omega$ for times close to where the diabatic energies cross, so by Eq.~(\ref{Eq:LZ}), the {\em adiabatic} transition probability is $P_{LZ} = 1 - \exp\left( \!  - \frac{\pi |\epsilon|^2}{4 \mu B_0 \hbar \omega} \!\right)$.  The dimensionless quantity 
\begin{equation} \label{eq:beta_new}
\beta \equiv  \frac{\mu B_0 \hbar \omega}{\pi |\epsilon|^2}
\end{equation}
is a measure of how adiabatic the transition is.  The pseudo-crossings of the energy eigenvalues 
\begin{equation} \label{eq:eig2} 
E_{\rm ad}(t)=\pm \sqrt{|\epsilon|^2+\mu^2 B_z(t)^2}+\frac13{\cal D} 
\end{equation}
occur at times $t = j\pi/2 \omega$, $j = 1, 3,5, \ldots$, as demonstrated in Fig.~\ref{Fig_1} for $0\leq t\leq \pi/\omega$. 
Here we only follow the dynamics through the first pseudo-crossing.  If we were to follow the dynamics over longer time periods, multiple pseudo-crossings would occur and  St\"uckelberg oscillations would play a role \cite{Huang_Du_11,Shevchenko_10}, but here we wish to focus on the adiabaticity in a single pseudo-crossing rather than coherent evolution between pseudo-crossings. 
\begin{figure}%[ht]
\centering
\includegraphics[width=0.7\linewidth]{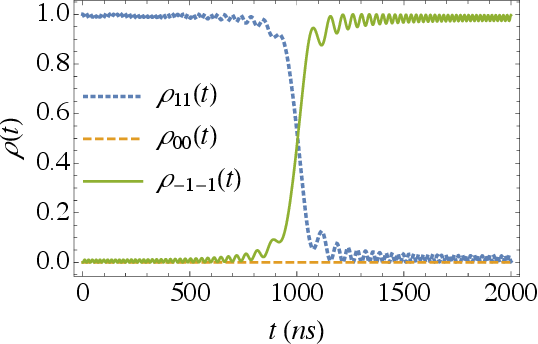}
\caption{Nearly adiabatic dynamics of the two-level system with the population initially in the $M_S = 1$ level.  The population is adiabatically transferred to the $M_S = -1$ level, and the $M_S = 0$ remains zero throughout the dynamics because with a magnetic field only along the $z$-axis, this component is completely decoupled if the evolution is unitary.  The parameters used in this calculation are as in Fig.~\ref{Fig_1} (adiabaticity parameter $\beta=0.26\ll 1$). }
\label{Fig_2}
\end{figure}
 
\begin{figure}%[ht]
\centering
\includegraphics[width=0.7\linewidth]{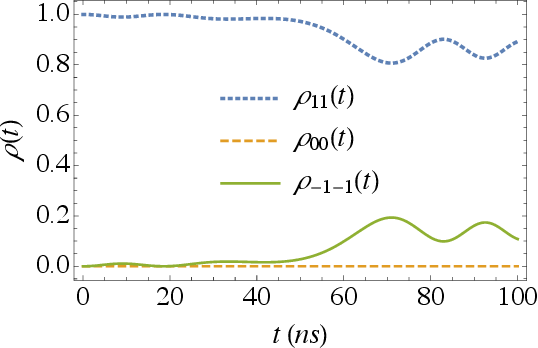}
\caption{Two-level non-adiabatic dynamics obtained with an angular frequency 20 times larger than that used in Figs.~\ref{Fig_2}, corresponding to an adiabaticity parameter $\beta=5.41\gg 1$. All the other parameters are identical. (To compare with Fig.~\ref{Fig_2}, scale the abscissa by a factor of 20).}
\label{Fig_3}
\end{figure}

\begin{figure}%[ht]
\centering
\includegraphics[width=0.7\linewidth]{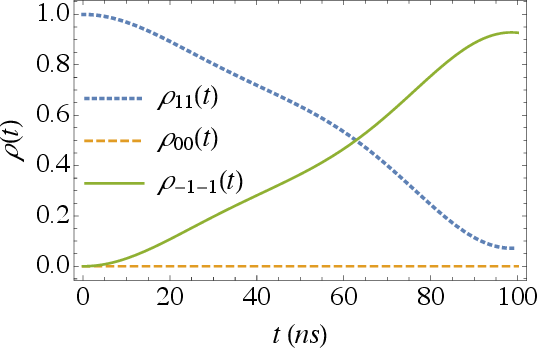}
\caption{Two-level non-adiabatic dynamics obtained with the same parameters used in Fig.~\ref{Fig_3}, but the magnetic field strength is reduced by a factor of 5, i.e., $\mu B_z = 2 \pi \times 4.8$ MHz (corresponding to a magnetic field of 2 Gauss). The adiabaticity parameter is $\beta=1.08\sim 1$, the slope of the eigenenergies versus time is slower than in Fig.~\ref{Fig_3} and the dynamics here are less non-adiabatic. }
\label{Fig_4}
\end{figure}
In the adiabatic limit where the sweeping frequency is small such that $\beta \ll 1$ the instantaneous state of the system is given by the adiabatic state
\begin{equation}
|\psi(t)\rangle_{\rm ad} = \frac{\epsilon|-1\rangle+[E_{\rm ad}-\mu B_z(t)]|+1\rangle}{\sqrt{(E_{\rm ad}-\mu B_z(t))^2+|\epsilon|^2}}, 
\end{equation}
which is dominated by the state $|+1\rangle$ at one side of the transition and by $|-1\rangle$ at the other side of the transition. This is demonstrated in Fig.~\ref{Fig_2}, where initially ($t=0$) the  population is in the state $|+1\rangle$, which almost completely coincides with one adiabatic state and ends up in the final time with the population almost completely in the $|-1\rangle$ state, which coincides with the same adiabatic eigenstate [the upper branch in Fig.~\ref{Fig_1} and Eq.~(\ref{eq:eig2})], corresponding to the magnetic moment aligned along the instantaneous axial magnetic field component. 
In the other limit where $\beta\gg 1$, the population of the $|M_S=\pm1\rangle$ states does not change during the sweeping, so that the final magnetic moment is aligned opposite to the axial magnetic field components. 
In the intermediate case of $\beta\sim 1$ a partial transition between the two states occurs, as demonstrated in Figs.~\ref{Fig_3} and \ref{Fig_4}.

\end{document}